\documentclass[%
 reprint,
 amsmath,amssymb,
 aps,
]{revtex4-1}

\usepackage{graphicx}
\usepackage{dcolumn}
\usepackage{bm}

\newcommand{\aap}{Astronomy \& Astrophysics}

\begin{document}

\title{Nuclear fission chain reaction in cooling white dwarf stars}

\author{C. J. Horowitz}\email{horowit@indiana.edu}
\affiliation{Center for Exploration of Energy and Matter and
                  Department of Physics, Indiana University,
                  Bloomington, IN 47405, USA}

\author{M. E. Caplan}
 \email{mecapl1@ilstu.edu}
\affiliation{
 Illinois State University, Department of Physics, Normal, IL 61790 
}%

\date{\today}

\begin{abstract}
The first solids that form as a white dwarf (WD) starts to crystallize are expected to be greatly enriched in actinides.  Previously [PRL {\bf 126}, 1311010] we found that these solids might support a nuclear fission chain reaction that could ignite carbon burning and provide a new Type Ia supernova (SN Ia) mechanism involving an {\it isolated} WD.  Here we explore this fission mechanism in more detail and calculate the final temperature and density after the chain reaction and discuss a number of open physics questions.

\end{abstract}

\maketitle

Material at white dwarf (WD) densities is ionized and crystallizes to form a Coulomb solid where the melting temperature of different chemical elements scales with atomic number $Z^{5/3}$ \cite{CaplanRMP}. 
Phase separation and crystallization can play important roles during the cooling of WD stars.  The Gaia space observatory has determined parallax distances to large numbers of galactic stars \cite{GaiaNoAuthors}, which allow for unprecedented modeling of WD and their evolution. Core crystallization, long predicted, is now resolved \cite{VanHorn1968,Tremblay2019}.  Phase separation or sedimentation of the neutron rich isotope $^{22}$Ne, in a C/O WD, could release significant gravitational energy and delay cooling \cite{Caplan_2020,Bildsten_2001,PhysRevE.82.066401,2020arXiv200713669B,Camisassa2020,Cheng_2019,bauer2020multigigayear}.

Pure uranium ($Z=92$) has a melting temperature 95 times higher than the melting temperature of C.  When a WD starts to crystallize, the first solids may be very strongly enriched in actinides, because these have the highest $Z$.           

In a previous paper \cite{PhysRevLett.126.131101} we found that these first solids could be so enriched that they support a fission chain reaction.  
Fission could possibly ignite carbon burning and produce a thermonuclear supernova (SN Ia).  These stellar explosions are important distance indicators in cosmology \cite{Abbott_2019,SN_cosmology,Sullivan2010}.  The exact SN Ia explosion mechanism is poorly understood but is thought to involve a WD interacting with a {\it binary} companion that is itself either a WD or a conventional star \cite{2012NewAR..56..122W,hillebrandt2013understanding,RUIZLAPUENTE201415}.  Presently there is tension between Hubble constant values determined from SN and in other ways \cite{Riess_2021,di_valentino_2021}.   This tension provides additional motivation to study SN Ia mechanisms and to try and understand how thermonuclear supernovae explode.        

\begin{figure}[tb]
\centering  
\includegraphics[width=0.35\textwidth]{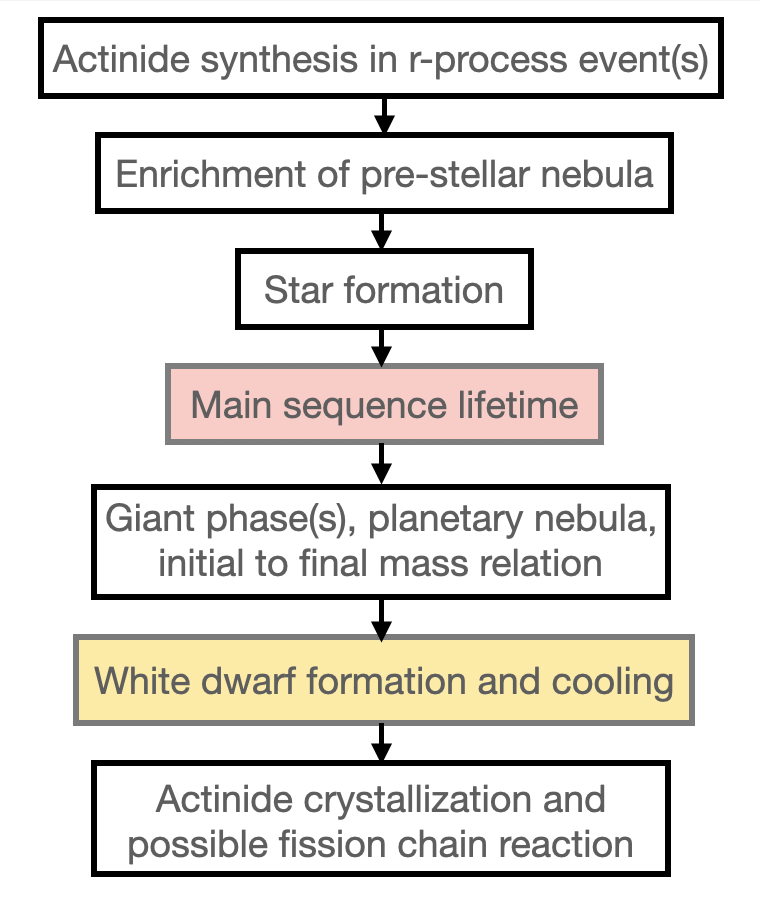}
\caption{\label{Fig2} Sequence from uranium production in an r-process nucleosynthesis event (top) to a fission chain reaction (bottom), see text.}	
\end{figure}

In the present paper we explore further the SN mechanism proposed in ref.  \cite{PhysRevLett.126.131101}  where fission in an {\it isolated} WD ignites carbon burning.  This mechanism involves the steps shown in Fig. \ref{Fig2}.
To begin, U must be produced in an r-process event and be incorporated into a pre-stellar nebula. During the main sequence the U remains mixed through the star, and following the giant phases and expulsion of a planetary nebula the resulting WD retains some of this U. As the WD cools some U may sink and enhance the core U number abundance. When the core cools to the actinide freezing temperature the actinides begin to crystallize, effectively forming precipitates, which may be so enhanced in U that they undergo a fission chain reaction which triggers a supernova. 

We discuss the composition of the first solids when uranium crystallizes, the criticality of this composition for a fission chain reaction, the configuration of the system and the size of the central region, the critical mass,  energy transport during fission  including losses from heat conduction, and the final temperature and density that will be reached.  We end by identifying and discussing a number of possibly open issues with this new SN mechanism as listed in Table \ref{Table4}.

\begin{table}[tb]
\caption{\label{Table4} Selected physics issues discussed in text.}
\begin{tabular*}{0.41\textwidth}{c c  } \hline \hline
Issue & Impacts \\ \hline
Enrichment of $^{235}$U & Criticality \\
Delay time & Enrichment \\
Mass of WD & Delay time\\
Amount of C/O in crystal & $n$ moderation, spectra \\
Amount of Pb in crystal & Heat capacity, final $T$ \\
Mass of central region (pit) & Total E released \\
T dependence of $\sigma_f(^{238}$U)& Fission E released \\ 
Heat conduction losses & Final $T$ and C ignition\\
CO vs ONe WD & Carbon ignition \\
\hline
\\
\end{tabular*}
\end{table}

{\it Initial enrichment of $^{235}$U during nucleosynthesis:}  We start with the initial enrichment $f^0_5$.  This is the fraction of U that is $^{235}$U following nucleosynthesis.  Actinides are very likely produced during the r-process where seed nuclei rapidly capture many neutrons \cite{Horowitz_2019,physicsreports_rprocess,2001AA...379.1113G,Truran1998}.  Recently, observations of the kilonova associated with the gravitational wave event GW170817 strongly suggest that neutron star mergers are an important r-process site.  Modern simulations of r-process nucleosynthesis in neutron rich ejecta during a merger find a ratio of $^{235}$U (and mother nuclei that will later decay into $^{235}$U) to $^{238}$U (and mother nuclei for $^{238}$U) near one.  This implies an initial enrichment $f_5^0\approx 50\%$. There are nuclear physics uncertainties for $f_5^0$ from for example the nuclear mass model used to describe heavy very neutron rich nuclei.  These uncertainties may be relatively modest because $^{235}$U and $^{238}$U are very near in mass.  Therefore changes in mass model may change the total U production more than the isotopic ratio.  Furthermore, new radioactive beam accelerators such as the Facility for Rare Isotope Beams (FRIB) can produce and study many of the neutron rich nuclei involved in the r-process and FRIB experiments may further reduce nuclear physics uncertainties.  

In addition to nuclear physics uncertainties, there are astrophysical uncertainties associated with conditions in the r-process site.  Many of these uncertainties are associated with the electron fraction $Y_e$ that governs the ratio of neutrons to protons in the r-process events ejecta.   One likely needs a moderately small $Y_e$ in order to significantly produce actinides.  Given this, the dependence of the isotopic ratio on $Y_e$ appears to be small \footnote{Erika Holman private communication}.  Therefore the astrophysical uncertainties in $f_5^0$ may also be modest, see for example \cite{10.1111/j.1365-2966.2012.21859.x}.  These uncertainties should be studied further in future work.  For now we assume r-process events synthesizes U with $f_5^0\approx 0.50$.

This fresh r-process material may then mix with older U that has a lower $f_5$ because some of the older U has had time to decay.   The solar system provides an example of how this might work.  Radioactive dating suggests the last event that contributed significant r-process material to the solar system occurred about 100 Myr before solar system formation \cite{Lugaro650}.  Material from this event and earlier production contributed to the total U inventory for the solar system.  The total enrichment at formation must have been $f_5\approx 0.25$ so that radioactive decay over the 4.6 Gyr lifetime of the solar system could result in the observed $f_5=0.7\%$ today.  Thus $f_5^0\approx 50\%$ was diluted to $f_5\approx 25\%$ by mixing with material from earlier r-process events.
 

{\it Delay time:} The delay time is the total time for all of the steps shown in Fig. \ref{Fig2} to occur and is important because U is radioactive. As $^{235}$U decays  $f_5(t)$ will decrease with time $t$ as,
\begin{equation}
f_5(t)=\Bigl[\bigl(\frac{1}{f_5^0}-1\bigr)\exp(\frac{t}{\tau})+1\Bigr]^{-1}\,
\end{equation}
with $\tau\approx 1.20$ Gyr.  If $f_5^0\approx 0.50$ after nucleosynthesis and one needs $f_5\approx 0.12$ or larger for criticality (see below), then $t$ must be less than 2.4 Gyr or one will have $f_5<0.12$ when U crystallizes so that  a chain reaction is no longer possible.  If the average enrichment from multiple r-process events, shortly after the last event, is only 0.25 then $t$ must be less than 1.1 Gyr in order for $f_5>0.12$ at the time of U crystallization.

Some contributions to the delay time include (1) the time between nucleosynthesis and star formation, (2) the main sequence lifetime of the star, and (3) the cooling time of the WD before U crystallization.   Both the main sequence lifetime and the WD cooling time depend strongly on mass and decrease rapidly for larger masses.   The main sequence lifetime of a star of mass $M_{ms}$ is $\approx 10(M_\odot/M_{ms})^{2.5}$ Gyr.  The star may then lose considerable mass before becoming a WD with mass $M_{wd}<M_{ms}$.   The cooling time of the WD depends on the crystallization temperature $T_c$ which we characterize with the Coulomb parameter $\Gamma\propto 1/T_c$, see after Eq. \ref{eq.Lambda}.  A 50/50 mixture of C/O crystallizes near $\Gamma\approx 200$ \cite{PhysRevLett.104.231101}.  Previously we found that U is expected to crystallize out at about twice the C/O temperature \cite{PhysRevLett.126.131101}.  Therefore, we calculate the cooling time of the WD for U crystallization $t_U$ as the time until $\Gamma\approx 100$ at the center of the star.  We note that this time is much less than the time until C/O crystallize $t_{C/O}$.  This is because the cooling rate depends on the star's luminosity and this decreases significantly as the star cools.  Therefore, it takes a WD more than three times as long to cool to $\Gamma=200$ as it does to $\Gamma=100$.    

\begin{figure}[tb]
\centering  
\includegraphics[width=0.48\textwidth]{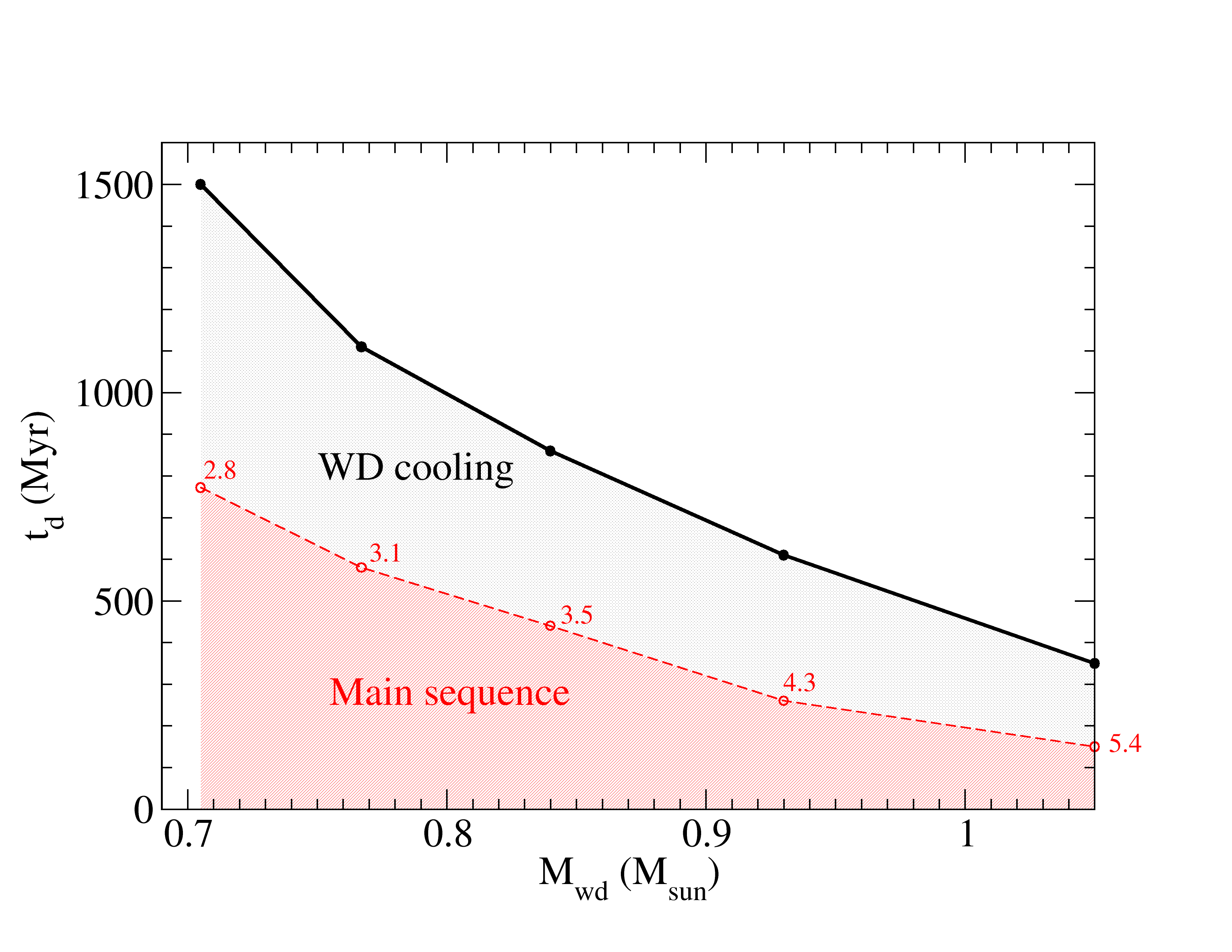}
\caption{\label{Fig3} Time delay $t_d$ from star formation until U crystallization (solid black line) versus WD mass.   
The red shaded region shows the main sequence lifetime and the red numbers indicate the original main sequence mass in M$_\odot$.  The black shaded region shows the WD cooling time until U crystallization. }	
\end{figure}

We list some of these times in Table \ref{Table5} and plot them in Fig. \ref{Fig3} for stars of different masses.  We use the initial final mass relation (IFMR) from ref. \cite{Cummings_2018} to relate $M_{wd}$ and $M_{ms}$ and we use the H poor evolutionary sequences from ref.~\cite{Camisassa_2017} to determine cooling times.  White dwarfs significantly less massive than 0.84 M$_\odot$ will have $t_{ms}+t_U\gg 860$ Myr and $f_5$ could be too small for the system to be critical.  White dwarfs more massive than 0.93 M$_\odot$ will have $t_{ms}+t_U<610$ Myr.  Therefore, we are primarily interested in massive WD with $M_{wd}$ larger than $\approx 0.84$ M$_\odot$.  Lower mass stars may have too long a time delay for a fission chain reaction.  

\begin{table}[tb]
\caption{\label{Table5} Main sequence mass $M_{ms}$ and lifetime $t_{ms}$ and WD mass $M_{wd}$ and WD cooling time for U crystallization $t_{U}$, see text.}
\begin{tabular*}{0.266\textwidth}{ccccc } 
$M_{ms}$ & $t_{ms}$ & $M_{wd}$ & $t_{U}$ & $t_{ms}+t_{U}$ \\ 
M$_\odot$ & Myr & M$_\odot$ & Myr & Myr \\ \hline
2.79 & 770 & 0.705 & 730  & 1500\\
3.12 & 580 & 0.767 & 530  & 1100 \\
3.50 & 440 & 0.837 & 420  & 860\\
4.33 & 260 & 0.934 & 350  & 610\\
5.41 & 150 & 1.050 & $\approx200$ & $\approx350$\\
\end{tabular*}
\end{table}

{\it Progenitor Masses:} The upper limit on $M_{wd}$ may be set by electron capture or by the composition of very massive WD.  Above a density of about $9.2\times 10^7$ g/cm$^3$ electron capture may remove $^{235}$U via $^{235}$U$(e,\nu_e)^{235}$Pa and likely prevent the actinide crystal from becoming critical.  This central density corresponds to a C/O WD with a mass $\approx 1.15$ M$_\odot$.  Interestingly, above $1.98\times 10^8$ g/cm$^3$ Pa may capture another electron to produce $^{235}$Th.  This nucleus with even $Z$ and odd $N$ may be fissile and could support a chain reaction.  Thus a very massive WD with $M_{wd}\approx 1.25$ M$_\odot$ may undergo $^{235}$Th fission.  

However, before these large masses are reached for electron capture, the main sequence star may have burned some carbon in its core to produce an O/Ne WD instead of a C/O WD.  It may be significantly harder for fission to ignite O burning instead of C burning, although this should be checked.  Therefore, the upper limit on the WD mass could be set by its composition.  White dwarfs more massive than about 1.05 M$_\odot$ may have O/Ne cores.  We conclude that our fission mechanism may be most straight froward for C/O WD with masses between $\approx 0.85$ M$_\odot$ and $\approx 1.05$ M$_\odot$.  These WD masses may correspond to main sequence masses between 3.5 M$_\odot$ and 5.4 M$_\odot$.  For an initial mass function $\propto M^{-2.3}$ we note that there are about 125\% as many stars between 3.5 and 5.4 $M_\odot$ compared to stars with masses above 8 M$_\odot$.  If a large fraction of these massive stars produce core collapse supernovae then there are very likely plenty of lower mass stars to produce SN Ia.  Note it could be that only a fraction of WD in this mass range have $f_5$ large enough for fission to ignite a SN.

{\it Composition of first solids:} We are interested in the very first solids to form as a WD just starts to crystallize.  
In the dense interior of a WD electrons form a degenerate Fermi gas.   As a result, conventional chemistry based on closed electron shells is replaced by the chemistry of a Coulomb plasma where the melting temperature of different elements scales with atomic number $Z^{5/3}$. As such, we expect them to be actinide-rich because these elements have the largest $Z$.

Ref. \cite{PhysRevLett.126.131101} predicted the composition of the first solid to include significant Pb, U and Th and also some C and O.   A simple linear mixing model for the free energies was used.  More recently we have performed molecular dynamics simulations that find two possible solid phases.  One contains a mixture of Pb, U and Th and no C and O and the other contains the heavy elements (Pb, U and Th) in a one-to-one ratio by number with a mixture of C and O.  We will present these MD simulations in a later publication.  For now we adopt the simpler composition, without any light C and O, as a model starting point, see Table \ref{Table1}.  Although Pb has a significantly smaller $Z=82$ than $Z=92$ for U, we still expect significant Pb in the first solids because the solar abundance of Pb is $\approx 100$ times larger than that for U.  We emphasize that uncertainties in the model composition of Table \ref{Table1} should be explored in future work.     

\begin{table}[tb]
\caption{\label{Table1} Composition of central pit (abundance by mass), electron fraction $Y_e^{pit}$, average charge $\bar Z$, and average mass $\bar A$.}
\begin{tabular*}{0.3035\textwidth}{c c c c c c} 
$x_U$  & $x_{Th}$ & $x_{Pb}$ & $Y_e^{pit}$& $\bar Z$ & $\bar A$\\  \hline
 0.287& 0.210 & 0.503 & 0.391& 86.26 & 220.78 \\
\end{tabular*}
\end{table}

{\it Criticality:} Following ref.~\cite{PhysRevLett.126.131101} we calculate the $^{235}$U abundance for criticality of an infinite system with the composition in Table~\ref{Table1}. We refer to the crystal as the pit (in analogy with a nuclear weapon).   The multiplication factor $k_\infty$ is equal to the number of fissions in one neutron generation over the fissions in the proceeding generation.  We write $k_\infty$ as the probability for fission of $^{235}$U times the number of neutrons released per fission $\nu$ over the probability of neutron absorption on various U and Th isotopes, 
\begin{equation}
k_\infty=\frac{\nu f_5 \sigma_f(^{235}{\rm U})}{f_5\sigma_a(^{235}{\rm U})+(1-f_5)\sigma_{n\gamma}(^{238}{\rm U})+N_{\rm Th} \sigma_{n\gamma}(^{232}{\rm Th})}\, .
\label{eq.k}
\end{equation}
This must be $\ge 1$ for a chain reaction.
Here $\sigma_f(^{235}{\rm U})$ is the fission cross section for $^{235}$U and $\sigma_a(^{235}{\rm U})$ is the sum of $\sigma_f(^{235}{\rm U})$ and $\sigma_{n,\gamma}$ for the $n,\gamma$ reaction.  Likewise $\sigma_{n\gamma}(^{238}$U) and $\sigma_{n\gamma}(^{232}$Th) are the $n,\gamma$ cross sections for $^{238}$U and $^{232}$Th.  Finally 
$N_{\rm Th}=238 x_{\rm Th}/(232 x_{\rm U})\approx 0.75$ is the number density of Th over the number density of U.  For simplicity we evaluate these cross sections for 1 MeV neutrons using the ENDF data set from Dec. 2011 that is available from the National Nuclear Data Center \cite{NNDC}, see Table \ref{Table2}.  We find $k_\infty\ge1$ for $f_5\ge0.12$.  Thus the $^{235}$U enrichment must be at least 12\% for the system to be critical.

\begin{table}[tb]
\caption{\label{Table2} Cross sections and fission neutron multiplicity $\nu$ for a neutron energy of one MeV \cite{NNDC}.}
\begin{tabular*}{0.416\textwidth}{c c c c c} 
$\sigma_f(^{235}$U)  & $\sigma_{n\gamma}(^{235}$U) & $\nu(^{235}$U) & $\sigma_{n\gamma}(^{238}$U) & $\sigma_{n\gamma}(^{232}$Th)\\  \hline
1.25 b& 0.105 b & 2.52 & 0.129 b & 0.143 b\\
\end{tabular*}
\end{table}

\begin{figure}[tb]
\centering  
\includegraphics[width=0.48\textwidth]{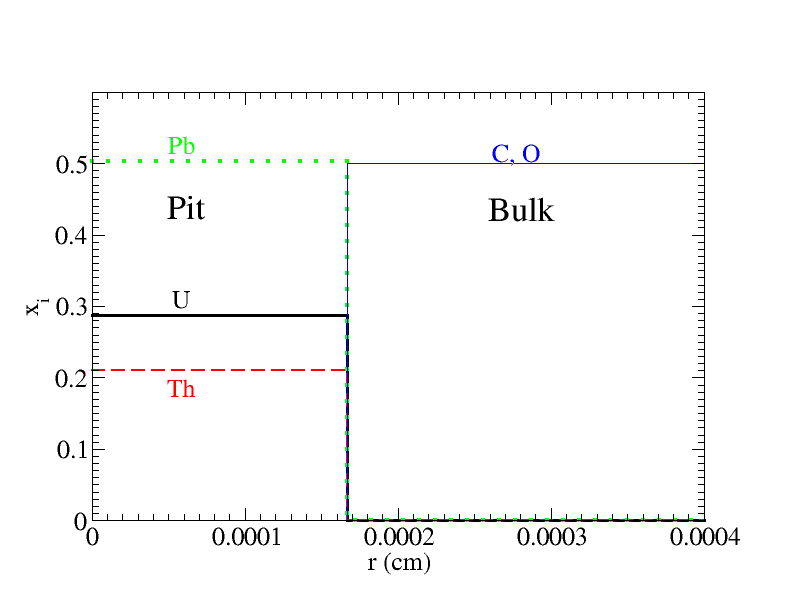}
\caption{\label{Fig1} Composition of central region or pit (left) and bulk liquid (right) vs distance $r$ from center of pit.}	
\end{figure}

{\it Configuration of system and size of central region or pit:} We expect a small seed crystal to grow by diffusion until a background neutron initiates a chain reaction.  The rate of crystal growth and the background neutron rate from spontaneous fission were discussed in ref. \cite{PhysRevLett.126.131101} where it was estimated that the crystal will grow to about 5 mg before a chain reaction is initiated.  We consider a C/O WD with a central density of about $2\times 10^8$ g/cm$^3$ and a spherical crystal of mass 5 mg and radius $r_p=1.67\times 10^{-4}$ cm.  Note the density of this crystal is slightly higher (by a factor of $0.5/Y_e^{pit}$), $2.56\times 10^8$ g/cm$^3$, because $Y_e^{pit}$ is less than 0.5.  The crystal is surrounded by a bulk liquid that is assumed to be a 50/50 mixture of C and O.   We expect fission, once initiated, to quickly melt the crystal.  This configuration is illustrated in Fig. \ref{Fig1}.

{\it Critical mass:} The critical radius $R_*$ of the system is the minimum size necessary for the system to be critical.  This is of order the neutron mean free path $l$.   If the system size is much smaller than $l$ too many neutrons will likely escape preventing a chain reaction.   A diffusion calculation predicts that $R_*$, using $k_\infty$ from Eq. \ref{eq.k}, is approximately \cite{primer,reed},
\begin{equation}
R_*=x\Bigl(\frac{l_t l_f/3}{k_\infty-1}\Bigr)^{1/2}\, .
\label{eq.R*}
\end{equation}
Here the value $x$ depends on properties of the ``tamper'' or bulk region in Fig. \ref{Fig1}, see below, and the critical mass is $M_*=4\pi R_*^3\rho/3$.  The mean free path for fission is $l_f^{-1}=n_Uf_5\sigma_f(^{235}$U) and the transport mean free path is $l_t^{-1}\approx (n_U+n_{Th}+n_{Pb})\sigma_t$ where $n_i$ are the number densities of the heavy elements and we assume equal transport cross sections of $\sigma_t\approx 4$ b.  If the transport mean free path in the tamper region of Fig. \ref{Fig1} is $l_t^{\rm tam}$, $x$ satisfies \cite{reed},
\begin{equation}
x\cot x = 1-\frac{l_t^{\rm tam}}{l_t}\, .
\label{eq.x}
\end{equation}
The C/O liquid surrounding the crystal provides a very good tamper that efficiently reflects neutrons back into the pit and significantly reduces the critical mass.    The transport cross section for scattering from C or O is comparable to that for scattering from the heavy nuclei however their number density is much higher.  As a simple approximation we assume the transport mean free path  scales with one over the ion density and the ion density is the electron density over the average charge $Z$.  Therefore  $l_t^{\rm tam}/l_t\approx 7/\bar Z$ where 7 is the average $Z$ of C/O and $\bar Z$ is from Table \ref{Table1} so that $x\approx 0.5$ from Eq. \ref{eq.x}. 
If $k_\infty\approx 1.1$ then $R_*\approx 3\times 10^{-6}$ cm and $M_*\approx 3\times 10^{-8}$ g \footnote{In ref. \cite{PhysRevLett.126.131101} we did not include the full effect of the tamper and got a larger $M_*$.}.  We note that $M_*$ is much less than the 5 mg mass of the pit and $R_*\ll r_p$.  Therefore we expect most neutrons to be absorbed in the pit and very few will escape out the surface.  

{\it Fission chain reaction:} We now consider a fission chain reaction in the pit that is initiated by a neutron from spontaneous fission.  We expect diffusion to be slow and viscosity to be small so that hydrodynamic evolution can be well approximated by a single equation describing energy conservation \cite{1992ApJ...396..649T},
\begin{equation}
\frac{\partial E(T)}{\partial t} + P \frac{ \partial(1/\rho_b)}{\partial t} = \frac{1}{\rho_b}\frac{\partial}{\partial x}\bigl(\sigma\frac{\partial T}{\partial x}\bigr) + \frac{\partial S}{\partial t}\, .
\label{eq.E}
\end{equation}
Here $E(T)$ is the energy per baryon of the  system. 
The temperature is $T$, $P$ is the pressure, $\rho_b$ is the baryon density, and $\sigma$ is the thermal conductivity.  Finally $S$ is the energy per baryon generated from fission.  Below we will argue that the thermal conductivity term is small compared to $\partial S/\partial t$.  If one neglects this term, Eq.~\ref{eq.E} can be integrated over time to yield,
\begin{equation}
E(T_f) - E(T_i) + P\bigl(\frac{1}{{\rho_b}_f} - \frac{1}{{\rho_b}_i}\bigr) \approx S\, .
\label{eq.Efi}
\end{equation}
Here $T_f$ is the final temperature after the fission chain reaction and $T_i\approx 0$ is the initial temperature.  Likewise ${\rho_b}_f$ is the final and ${\rho_b}_i$ the initial baryon density and we note that the system is at constant pressure.  

The energy per baryon of a low temperature very relativistic electron gas and ion system is
\begin{equation}
E(T)\approx Y_e^{pit}\bigl[\frac{3}{4}\epsilon_F^0+\frac{\pi^2}{2}\frac{(k_bT)^2}{\epsilon_F^0}\bigr] +\frac{3}{2\bar A}k_bT+ O(k_bT)^4\, ,
\label{eq.ET}
\end{equation}
with Fermi energy $\epsilon_F^0=(3\pi^2n_e)^{1/3}$ for initial electron density $n_e=Y_e^{pit}{\rho_b}_i$.  We neglect the electron mass in $\epsilon_F^0$. Finally $\bar A$ is the average mass number of the heavy nuclei.  This term in Eq. \ref{eq.ET} describes the heat capacity of the ions and is negligible compared to the first term \footnote{In Ref. ~\cite{PhysRevLett.126.131101}  we assumed the heat capacity was dominated by ions and neglected the electron gas contribution.  This oversight leads to an over estimation of the final temperature.}.  The pressure at low temperatures is,
\begin{equation}
P\approx\frac{{\epsilon_F}^4}{12\pi^2}+\frac{{\epsilon_F}^2(k_bT)^2}{18}+ O(k_bT)^4\, .
\label{eq.P}
\end{equation}
Setting $P$ to a constant allows one to solve for $\epsilon_F$ and the density $\rho$ as a function of temperature.  The result is
\begin{equation}
\rho(T_f)\approx \frac{\rho(T_i\approx 0)}{1+\frac{\pi^2}{2}\bigl(\frac{k_bT_f}{\epsilon_F^0}\bigr)^2}\, .
\label{eq.rho}
\end{equation}
For typical values $\epsilon_F^0\approx2$ MeV and $k_bT_f\approx0.5$ MeV one has $\rho(T_f)\approx 0.75 \rho(T_i=0)$.  This modest 25\% decrease in density is much too small to stop the fission chain reaction by allowing more neutrons to escape.  Given the small critical mass, it would take a decrease in density by a factor of $\approx 400$ to stop the chain reaction.  This is an important result that implies the fission reaction, once started, will likely continue to burn until so much U is fissioned that the system is no longer critical.

{\it Multiple neutron capture:}  We now discuss the fraction of U and Th nuclei that may fission.  Our system may behave in ways unlike conventional nuclear reactors or terrestrial atomic bombs.  In a reactor there is time for beta decay so that a reactor can breed $^{239}$Pu by neutron capture on $^{238}$U.  A conventional atomic bomb disassembles so rapidly that there may be limited time for heavy nuclei to capture two or more neutrons.  In our system $^{236}$U, $^{239}$U, and $^{233}$Th nuclei, formed by neutron capture, can capture additional neutrons.  Each of these nuclei has a significant fission cross section, although not as large as for $^{235}$U, see Table \ref{Table3}. Fission of these nuclei will contribute further heating and neutrons to the chain reaction.

\begin{table}[htb]
\caption{\label{Table3} Fission cross sections for additional neutron capture nuclei (at 1 MeV) \cite{NNDC}.}
\begin{tabular*}{0.18\textwidth}{c c c } 
$^{236}$U  & $^{239}$U & $^{233}$Th \\  \hline
0.36 b & 0.39 b & 0.095 b\\
\end{tabular*}
\end{table}

{\it Thermal effects:}  We expect fission cross sections for $^{238}$U and $^{232}$Th to significantly increase, as the temperature increases, because of thermal effects. Zhu and Pei calculate that the spontaneous fission half life of $^{240}$Pu decreases by 12 orders of magnitude as the temperature is increased from 0 to 0.1 MeV  \cite{PhysRevC.94.024329}.  We  expect significant increases in neutron induced fission cross sections with similar modest increases in temperature.  This should be verified with further theoretical work.  We expect a large fraction of both U and Th nuclei to eventually fission because of the chance to capture multiple neutrons and because of the increase in fission cross section with increasing temperature.

\begin{figure}[tb]
\centering  
\includegraphics[width=0.48\textwidth]{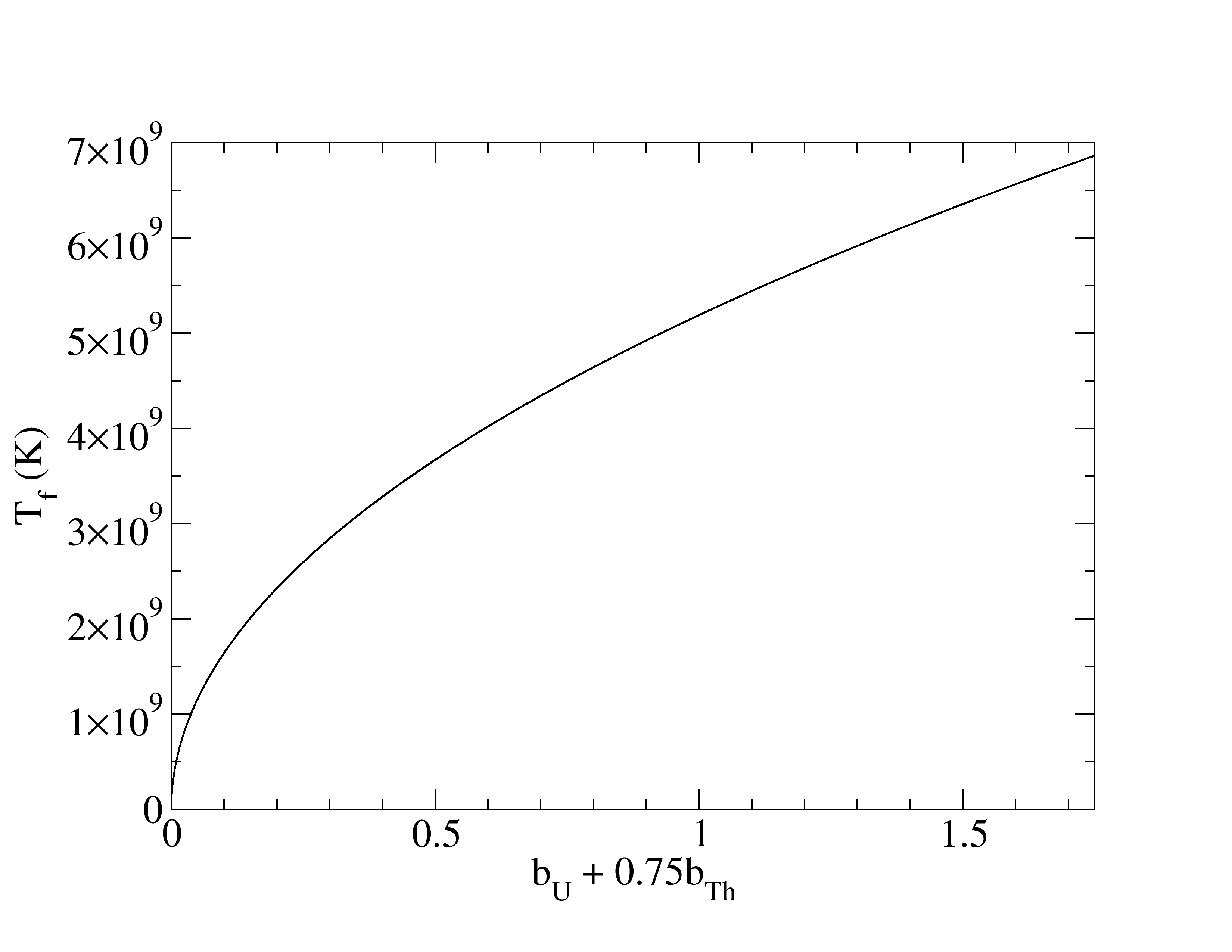}
\caption{\label{Fig4} Final temperature versus fraction of uranium $b_U$ or thorium $b_{Th}$ that fissions, see Eq. \ref{eq.Tf}. The factor of 0.75 reflects the lower Th number density.}	
\end{figure}

We write the total energy released in fission (per baryon) as,
\begin{equation}
S\approx 200 {\rm MeV}(\frac{x_U}{238}b_U+\frac{x_{Th}}{232}b_{Th})\, .
\end{equation}
Here $b_U$ is the fraction of all U nuclei ($^{235}$U, $^{239}$U, etc.) that eventually fission and $b_{Th}$ is the fraction of Th nuclei that fission.
Using Eqs.~\ref{eq.ET}, \ref{eq.P}, \ref{eq.rho} in Eq. \ref{eq.Efi}, we have the simple result that the integral of the heat capacity at constant pressure is equal to the energy released from fission, $5Y_e^{pit}\pi^2(k_bT_f)^2/(8\epsilon_F^0)=S$.  Solving for $T_f$ gives,
\begin{equation}
k_bT_f=\Bigl(\frac{8\epsilon_F^0 S}{5\pi^2 Y_e^{pit}}\Bigr)^{1/2}\, .
\label{eq.Tf}
\end{equation}
This is plotted in Fig. \ref{Fig4}.  For $\epsilon_F^0\approx 2$ MeV and the composition in Table \ref{Table1}, the final temperature is $T_f=6.9\times 10^9$ K if $b_U=b_{Th}=1$.  If only the U fissions we have $T_f=5.2\times 10^9$ K for $b_U=1$, $b_{Th}=0$.

{\it Heat conduction losses:} We neglected the heat conduction term involving $\partial/\partial x(\sigma\partial T / \partial x)$ compared to $\partial S/\partial t$ in Eq. \ref{eq.E}.  We obtain a first estimate of heat conduction by assuming the temperature varies over a length scale equal to the size of the system in Fig. \ref{Fig1} so that $\partial T/\partial r$ is of order $T/r_p$.  The thermal conductivity $\sigma$ is dominated by conduction from the degenerate electrons and the electron mean free path is limited by electron ion scattering.  We use the simple form in ref. \cite{1980SvA....24..303Y} where,
\begin{equation}
\sigma=\frac{\pi^3k_b^2Tn_e}{4\alpha^2\epsilon_F^2\bar Z\Lambda_{ei}}\, .
\end{equation}
Here the Coulomb logarithm describing electron ion scattering is
\begin{equation}
\Lambda_{ei}\approx \ln\Bigl[\bigl(\frac{2}{3}\pi\bar Z\bigr)^\frac{1}{3}\bigl(1.5+\frac{3}{\Gamma}\bigr)^\frac{1}{2}\Bigr] -\frac{1}{2}\, .
\label{eq.Lambda}
\end{equation}
The Coulomb parameter for our multicomponent plasma is $\Gamma=\langle Z^\frac{5}{3}\rangle e^2/(a_ek_bT)\approx 30.6$ at $k_bT=0.5$ MeV.  Here the electron sphere radius is $a_e=[3/(4\pi n_e)]^{1/3}$.  We note that the thermal conductivity is reduced by the high average charge $\bar Z$ of our system, see Table \ref{Table1}.  We have at $k_bT=0.5$ MeV and $r_p=1.7\times 10^{-4}$ cm,
\begin{equation}
\frac{1}{\rho_b}\frac{\partial}{\partial x}\sigma\frac{\partial T}{\partial x} \approx \frac{\sigma T}{\rho_br_p^2}\approx 6.0\times 10^8 {\ \rm MeV/s}\, .
\label{eq.sigma}
\end{equation}
We compare this to the rate of fission heating that we approximate as
\begin{equation}
\frac{d S}{dt} \approx \bigl(\frac{ 200 {\rm MeV}}{238}\bigr)\frac{ b_1}{\tau}\approx 3.5\times 10^{11} {\ \rm MeV/s}\, .
\label{eq.Sdot}
\end{equation}
Here the neutron fission mean free path over velocity is $\tau=l_f/v\approx 2.4\times 10^{-14}$ s and $b_1$ is the maximum fraction of U that fissions in any one neutron generation.  The fission rate rises exponentially during the chain reaction as the number of neutrons increases very rapidly.  What is most important is the comparison of the fission rate to the thermal conductivity loses when the fission rate is very high and most of the energy is being released.  We expect the fission rate to be very high near the end of the chain reaction, just as the system starts to run out fuel (nuclei to fission).  We somewhat arbitrarily consider $b_1\approx 0.01$.  With this choice the fission energy release in Eq. \ref{eq.Sdot} is 1,000 times larger than the heat conduction rate in Eq. \ref{eq.sigma}.  This justifies our neglect of heat conduction loses in Eq. \ref{eq.Tf}.

{\it Carbon ignition:} Timmes and Woosley have explored the conditions necessary to ignite C burning via a deflagration \cite{1992ApJ...396..649T}.  According to their Fig.~6, a trigger mass of $M^*=5$ mg needs to be heated above $\approx 5\times 10^9$ K for carbon ignition.  If a significant fraction of the U and or Th fissions then this temperature may be reached. Note that the pit is small and there is only a small amount of U in the star.  Therefore fission must ignite carbon fusion in order to provide enough energy for a SN.  After carbon ignition, the deflagration could possibly turn into a detonation \cite{Poludnenkoeaau7365}.  We conclude, it is possible that a fission chain reaction could ignite a thermonuclear supernova (SN).  However there are a number of possibly open issues and we discuss some of them now. 


{\it Importance of the composition of the first solids:}  We emphasize that the composition of the first solids is very important.  If more Pb is present the heat capacity would increase and the final temperature decrease.  This could possibly prevent C ignition.  If significant C and O are present, they would act as a moderator and reduce neutron energies.  This could prevent the system from becoming critical or slow down the fission chain reaction and allow more losses from heat conduction.  Finally C, if present in the crystal, could provide additional fusion energy.  This C should be relatively easy to fuse because of very strong screening from the high $Z$ elements. 

{\it Future work}  should be carried out in many areas.  For example, one should study further the phase diagram of multicomponent plasmas that include very high $Z$ elements to better constrain the composition of the first solids.  The temperature dependence of the neutron induced fission cross section on $^{238}$U should be calculated and one should study the total energy released from a fission chain reaction.  Hydrodynamic simulations should be performed to see if this fission reaction can ignite carbon burning, and if so will this lead to a SN Ia. Finally one should study U enrichment in r-process nucleosynthesis and if this enrichment will be high enough for criticality. 

{\it Multiple SN Ia mechanisms and cosmological bias:}  There may be more than one way to produce a thermonuclear supernova.  For example some SN with shorter delay times could be due to fission in a sub Chandrasekhar mass WD while SN with longer delay times could be from more traditional single degenerate or double degenerate mechanisms.  Furthermore, the relative contribution of different mechanisms could change with metallicity.  For example earlier in the history of a galaxy there may be less ``old'' $^{238}$U from earlier r-process events.  As a result the average U enrichment of a pre-stellar cloud could be higher and closer to the $\approx 50$\% enrichment of the latest r-process event.   The higher enrichment may allow lower mass stars, with longer delay times, to explode via fission ignition.  These lower mass stars will likely produce SN with less nickel and lower luminosities.   If this change in luminosity is not perfectly corrected by an empirical correlation with for example a light curve decay time, a cosmological bias could result.  

Presently there is tension between Hubble constant values determined from SN and in other ways \cite{Riess_2021,di_valentino_2021}.  It may be premature to speculate now on the size of a possible cosmological bias from our fission mechanism.  Nevertheless, the Hubble tension provides strong motivation to explore this and other SN Ia mechanisms and to try and better understand how thermonuclear supernovae explode.

{\it Conclusions:} The first solids that form as a white dwarf (WD) starts to crystallize are greatly enriched in actinides because of their large charges.  We estimate that these first solids could be so enriched in actinides that they may support a fission chain reaction.  This reaction could ignite carbon burning and lead to the explosion of an isolated WD in a thermonuclear supernova.

We thank Erika Holmbeck for sharing preliminary results on uranium synthesis in the r-process and for many comments on the manuscript.  We thank Rolfe Petschek for discussions on the structure of high $Z$ crystals. We thank Cameron Reed for pointing out that the C/O liquid is an excellent tamper.   We thank Ezra Booker, Alex Deibel, Witek Nazarewicz, Tomasz Plewa, and Rebecca Surman for helpful discussions.  This research was supported in part by the US Department of Energy Office of Science Office of Nuclear Physics grants DE-FG02-87ER40365 and DE-SC0018083 (NUCLEI SCIDAC).

\bibliography{apsbib}

\end{document}